# Venus as a Laboratory for Exoplanetary Science

*Stephen R. Kane[1], Giada Arney[2], David Crisp[3], Shawn Domagal-Goldman[2], Lori S. Glaze[2], Colin Goldblatt[4], David Grinspoon[5], James W. Head[6], Adrian Lenardic[7], Cayman Unterborn[8], Michael J. Way[9], Kevin J. Zahnle[10]*

[1] *University of California, Riverside, CA, USA*
[2] *NASA GSFC, Greenbelt, MD, USA*
[3] *JPL, Pasadena, CA, USA*
[4] *University of Victoria, Canada*
[5] *Planetary Science Institute, Tucson, AZ, USA*
[6] *Brown University, Providence, RI, USA*
[7] *Rice University, Houston, TX, USA*
[8] *Arizona State University, Tempe, AZ, USA*
[9] *NASA GISS, New York, NY, USA*
[10] *NASA Ames Research Center, Moffett Field, CA, USA*

Corresponding author: Stephen R. Kane (skane@ucr.edu)

Key points:
1. The characterization of terrestrial exoplanets, including interior structure and atmospheres, is becoming a primary focus of exoplanetary science.
2. The boundaries of habitability are best understood through the study of the extreme environments present on Earth and Venus.
3. There are many outstanding questions regarding Venus that are critical to answer in order to better constrain models for exoplanets.

# Abstract

The current goals of the astrobiology community are focused on developing a framework for the detection of biosignatures, or evidence thereof, on objects inside and outside of our solar system. A fundamental aspect of understanding the limits of habitable environments (surface liquid water) and detectable signatures thereof is the study of where the boundaries of such environments can occur. Such studies provide the basis for understanding how a once inhabitable planet might come to be uninhabitable. The archetype of such a planet is arguably Earth's sibling planet, Venus. Given the need to define the conditions that can rule out bio-related signatures of exoplanets, Venus provides a unique opportunity to explore the processes that led to a completely uninhabitable environment by our current definition of the term. Here we review the current state of knowledge regarding Venus, particularly in the context of remote-sensing techniques that are being or will be employed in the search for and characterization of exoplanets. We discuss candidate Venus analogs identified by the *Kepler* and *TESS* exoplanet missions and provide an update to exoplanet demographics that can be placed in the potential runaway greenhouse regime where Venus analogs are thought to reside. We list several major outstanding questions regarding the Venus environment and the relevance of those questions to understanding the atmospheres and interior structure of exoplanets. Finally, we outline the path towards a deeper analysis of our sibling planet and the synergy to exoplanetary science.

## 1. Introduction

The new era of exoplanet research provides a basis to place the terrestrial planets of our planetary system into a much broader context and explore a wide range of potential variability through comparative planetary system research. One of the most compelling questions in comparative planetology of our Solar System is the origin and evolution of life (astrobiology): when, where, how and under what conditions did life arise, and what environments encourage its evolution or cause its extinction? The prime focus of astrobiology research is the search for life elsewhere in the universe, and this search proceeds with the pragmatic methodology of looking for liquid water and Earth-like conditions. In our solar system, Venus is the most Earth-like planet, yet at some point in planetary history there was likely a bifurcation between the two: Earth has been continually habitable since the end-Hadean, whereas Venus became uninhabitable at some point in its past. Indeed, Venus may be the type-planet for a world that has transitioned from habitable and Earth-like conditions through the inner edge of the Habitable Zone (HZ); thus it provides a natural laboratory to study the evolution of habitability. Exoplanet detection methods are becoming increasingly sensitive to terrestrial planets, resulting in a much-needed collaboration between the exoplanetary science and planetary science communities to leverage the terrestrial body data within the solar system. In fact, the dependence of exoplanetary science on solar system studies runs deep, and influences all aspects of exoplanetary data, from orbits and formation, to atmospheres and interiors.

A critical aspect of exoplanetary science to keep in mind is that, unlike the solar system, in situ data for exoplanet surface environments will not be obtained in the foreseeable future, and thus exoplanet environments may only be characterized indirectly from other measurables, such as planetary mass, radius, orbital information, and atmospheric composition. Inferences about those environments in turn are derived from detailed models constructed using the direct measurables obtained from observations of and missions to solar system bodies (Fuji et al. 2014; Madden & Kaltenegger 2018). Thus, even as we struggle to understand the fundamental properties of terrestrial objects within the solar system, the task of characterizing the surface environments of Earth-sized planets around other stars will remain ever moreso inaccessible. If we are then to seek to understand the habitability of planets similar to the Earth, proper understanding of the boundaries of the HZ are necessary, exploring both habitable and uninhabitable environments. Furthermore, current and near-future exoplanet detection missions are biased towards close-in planets (i.e., those with relatively short orbital periods), so the most suitable targets for the *James Webb Space Telescope (JWST)* are more likely to be Venus-like planets than Earth-like planets. Thus, the ongoing efforts to further study and understand the evolution of Venus' atmosphere and surface, including its present state, provides critical information that complements the interpretation of these exoplanet observations. Here, we review the current state of knowledge regarding Venus in the context of habitability and the potential of past temperate conditions. We further discuss the relevance of Venus to the study of terrestrial exoplanets, as well as current and future exoplanet missions, and the primary outstanding questions on Venus, the answers of which will greatly inform our understanding of terrestrial planetary evolution and habitability in general.

## 2. The Current Venus Environment

Venus can be considered an "Earth-like" planet, because it has a similar size and possibly the same bulk composition (Zharkov, 1983) as its slightly larger neighbor. However, it has a 92 bar atmosphere consisting 96.5% $CO_2$ and 3.5% $N_2$, and a surface temperature of 735 K. Shown in the top panel of Figure 1 is an early topographical map of the Venusian surface produced via radar mapping by the Pioneer Venus orbiter, exhibiting the dominant highlands of Ishtar Terra in the north and Aphrodite Terra near the equator. The bottom panel of Figure 1 shows a map produced by the Magellan spacecraft with an improved spatial resolution (less than 100m). Venus' atmospheric composition and pressure is well explained by a runaway greenhouse having occurred in the past (Walker 1975), when insolation exceeded the limit on outgoing thermal radiation from a moist atmosphere (Komabayashi 1967; Ingersoll 1969; Nakajima et al. 1992; Goldblatt & Watson 2012; Goldblatt et al. 2013), which led to the evaporation of oceans presumed to have been present. It is unclear whether the oceans condensed, then later evaporated, or never condensed after accretion (Hamano et al. 2013). In either case, water loss by hydrogen escape followed, evident by the substantially high D/H ratio found for Venus relative to Earth (Donahue 1982). Complete water loss would take a few hundred million years (Watson et al. 1981), but may have been throttled by oxygen accumulation (Wordsworth & Pierrehumbert 2014). Notably, massive water loss during a runaway greenhouse has been suggested as a means to produce substantial $O_2$ in exoplanet atmospheres (Luger & Barnes

2015), but Venus serves as a counter-example to this concept. The hydration of surface rocks (Matsui & Abe 1986a) or top-of-atmosphere loss processes (Chassefière 1997; Collinson et al. 2016) are potential sinks for water. Thus, Venus is an ideal laboratory to test hypotheses of abiotic oxygen loss processes. Tracers for water loss processes on Venus are discussed in detail in Section 3.

Cloud-top variations in $SO_2$ have been documented over several decades from *Pioneer Venus* to *Venus Express* (Marcq et al. 2012). These observations imply a long-term atmospheric cycling mechanism, or possibly injections via volcanism. Recently, the Visible and Infrared Thermal Imaging Spectrometer (VIRTIS) identified nine emissivity anomalies attributed to compositional differences as sites of potentially recent volcanism (Smrekar et al. 2010, hereafter S2010). These anomalies have purported associated lava flows that are estimated to have a maximum age of 2.5 million years. However, it is more likely that their age is 250,000 years old or less (S2010) based on the expected weathering rates of freshly emplaced basalts. Furthermore, *Magellan* gravity data indicates that the emissivity anomalies are associated with regions of thin, elastic lithosphere , strengthening the volcanism interpretation. A new analysis of *Venus Express*' Venus Monitoring Camera (VMC) data in 2015 uncovered dditional evidence for active volcanism on Venus. Four temporally variable surface hotspots were discovered at the Ganiki Chasma rift zone, located near volcanoes Ozza and Maat Montes (Shalygin et al., 2015), suggestive of recent or ongoing volcanic activity. However, correct interpretation of these types of observations from above the cloud layer is challenging. The extent of scattering radiation from Venus's surface escaping through the cloud deck is about 100 $km^2$, so smaller regions of increased thermal emission are not accurately resolved.

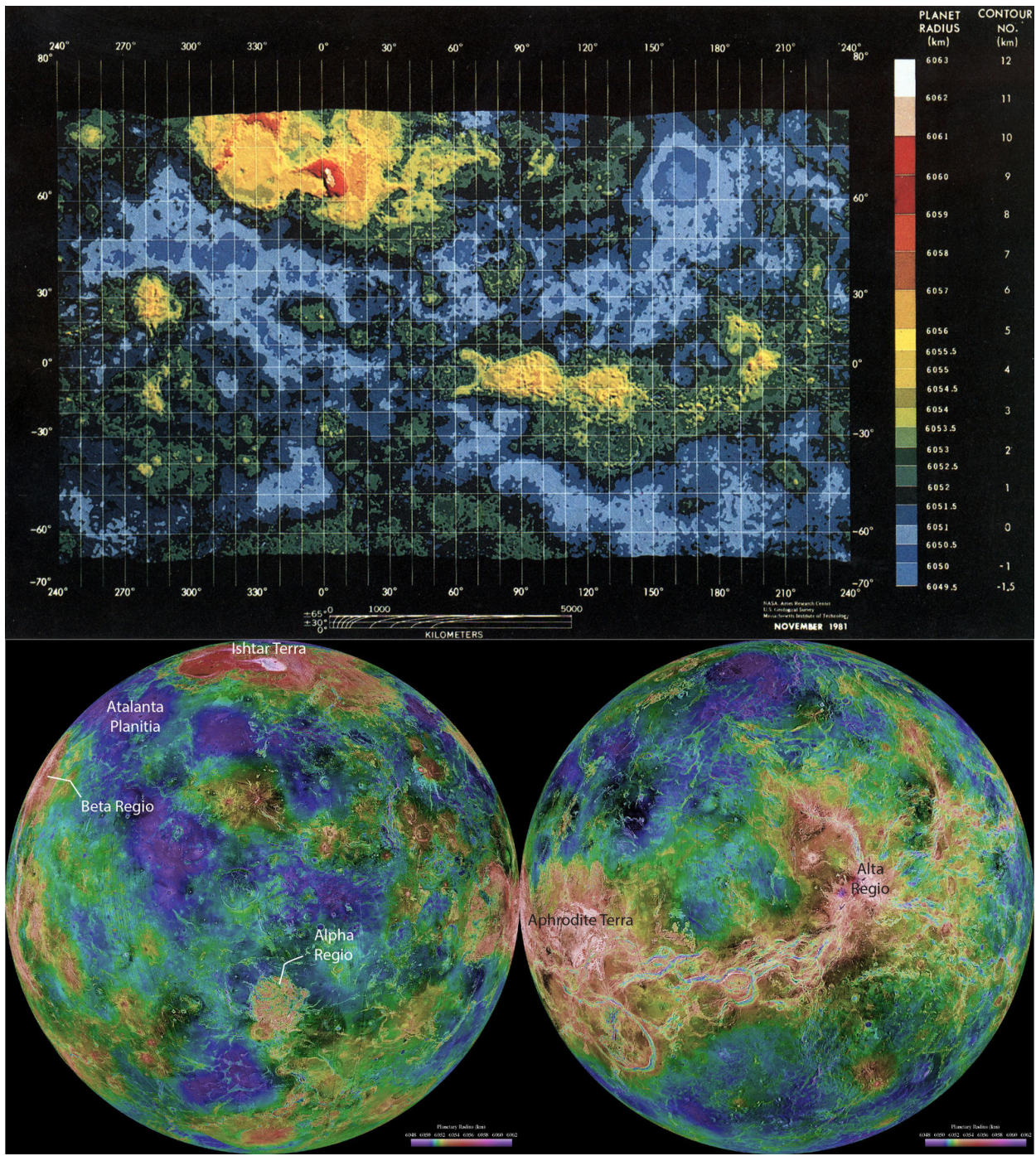

*Figure 1: Topographical map of the Venusian surface based on observation from Pioneer Venus orbiter (top) and the Magellan spacecraft (bottom) Credit: NASA Ames Research Center, US Geological Survey, Massachusetts Institute of Technology, and the NASA Space Science Data Coordinated Archive.*

Recently, data from the JAXA *Akatsuki* spacecraft have shown evidence of striking UV-bright stationary wave in the Venus upper atmosphere (Fukuhara et al. 2017). The center of this feature appears to be located above the Aphrodite Terra highland region and may be a stationary gravity wave caused by deep atmosphere winds flowing over the elevated terrain. Similar features were also seen in the 1980s by the Soviet VEGA balloons (Young et al. 1987) and in analyses of *Venus Express* data, in which such features are again associated with topographic highs (Peralta et al. 2017). In addition to topography, latitude and diurnal effects appear to influence these waves, suggesting a complex interplay between atmospheric dynamics and solar heating (Kouyama et al. 2017). Such interactions between planetary topography and atmospheric dynamics may be critical in the correct interpretation of exoplanet atmosphere data, and may be used to more robustly infer exoplanet surface conditions.

## 3. Diagnosing the History of Water Loss on Venus

Today Venus has just 0.001% of an Earth-ocean equivalent volume of water in its atmosphere and an unknown but probably small amount of $H_2O$ stored in its mantle. By contrast, Earth has an ocean of water at the surface and probably a corresponding water volume or two in its mantle (Korenaga 2017). It is possible that Venus initially formed with much less water than Earth, but because water-bearing planetesimals are nearly as likely to hit Venus as Earth (Wetherill 1981), it is likely that Venus accreted with an amount of water commensurate with its reservoirs of $CO_2$ and $N_2$, both of which are similar to Earth.

The runaway greenhouse is the accepted explanation for how Venus lost its water (Ingersoll 1969). Under this scenario, the atmospheric cold-trap for water disappears with increasing incident solar radiation (insolation). With water vapor abundant in the atmosphere, rapid hydrogen escape is possible at a rate controlled mostly by the Sun's extreme ultraviolet (EUV) radiation, which was greater when the Sun was younger. At very early times the hydrogen from an ocean of water could have been lost in less than 10 million years, whereas with today's mature Sun it would take a billion years (Abe et al. 2011). Assuming a similar initial water inventory to Earth, the classic confirmation of massive hydrogen escape from Venus is the extraordinarily high D/H ratio measured by *Pioneer Venus* in a trapped droplet of sulfuric acid (Donahue & Pollack 1983; Donahue & Russell 1997; Donahue 1999), indicating that at least 99% of the initial inventory has been lost since planetary formation. It was proposed by Grinspoon & Lewis (1988) that the observed Venus D/H ratio is consistent with a steady state of water loss where water is regularly provided by cometary impacts. However more recent atmospheric erosion simulations by Kulikov et al. (2006) predict excessive hydrogen escape that is indicative of significant water loss.

One proposed scenario for when Venus lost its water inventory is that Venus was always too hot from high insolation for water to condense (Matsui & Abe 1986b; Gillman et al. 2009, Hamano et al. 2013). Hydrogen escape would then have taken place early and rapidly (Zahnle et al. 1988), consequently driving off many other atmospheric constituents and imprinting the remaining gases with telltale mass fractionations (Hunten et al. 1987; Pepin 1991). Alternatively, the onset of the runaway greenhouse effect may have been delayed by the low luminosity of the young Sun (Gillman et al. 2009; Abe et al. 2011; Hamano et al. 2013). Other constraints, such as the limited atmospheric inventory of measured radiogenic $^{40}Ar$, which implies an early shut off of Earth-like degassing, argue strongly that even under this scenario the loss of most of the planet's hydrogen took place no later than ~3.5 Gya (Turcotte & Schubert 1988). Thus, to satisfy measured data for Venus, the second scenario must include a means by which leftover oxygen was sunk into the mantle, which may conflict with the apparently mildly reduced state of Venus's atmosphere and surface (Fegley 2003).

Other noble gas abundances provide various indicators of water loss. Currently helium does not appear to be escaping at a significant rate from Venus (Donahue & Russell 1997). Thus it may be possible to use the radiogenic $^4$He inventory, created by Th and U decay, to place constraints on how long Venus has been in its present state. *Pioneer Venus* measured $^4$He at altitudes above the mixed atmosphere (Donahue & Pollack 1983). When extrapolated to the lower atmosphere, the upper bound is consistent with little or no $^4$He escape over the past 3.5 billion years, implying a long-dormant Venus. Krypton is relatively massive and difficult to ionize, and hence it is very difficult for this gas to escape by any process other than impact erosion. There is no evidence of fractionating Kr escape from either Earth or Mars, but detection of strongly mass-fractionated Kr on Venus would provide evidence of an early $H_2$-dominated atmosphere (Pepin 1991).

Non-radiogenic xenon on Earth and Mars is very strongly mass fractionated and depleted, implying that Xe must have escaped these planets. To date, arguably the only tractable model for Xe escape proposes xenon ion being dragged to space by ions in a hydrogen-rich planetary wind (Zahnle et al. 2019). Under this scenario, Xe escapes at relatively modest levels of solar EUV and hydrogen escape fluxes, and on Venus this mechanism can potentially probe times as recent as 1 Gya. Xenon also has several radiogenic isotopes that are daughters of short-lived radioactive nuclei. Strange anomalies in these isotopes, such as unexpected abundances, would require processes taking place very early in Venus's history that alter the relative abundance of Xe in the Venus atmosphere, and hence are most useful in probing the first scenario under which water loss takes place very early in Venus's history. If non-radiogenic $^{36}$Ar/$^{38}$Ar in the Venus atmosphere turns out to be low ( as is the case for Mars), this finding would provide strong evidence that early hydrogen escape was efficient and that $CO_2$ was also subject to escape (Zahnle et al. 1990). If the Venusian $^{36}$Ar/$^{38}$Ar ratio is that of the solar ratio then Ar has not escaped and neither has hydrogen nor $CO_2$ at high rates. Neon may escape if the hydrodynamic hydrogen escape rate is sufficiently high (Ozima & Zahnle 1993). If $^{20}$Ne/$^{22}$Ne is found for Venus to be the same as the solar ratio, then neon did not escape. Lack of neon escape could imply that water loss from Venus was delayed, as in the second scenario described above. Highly fractionated neon, i.e., a – $^{20}$Ne/$^{22}$Ne ratio substantially less than that of the ratio for Earth, would be strong evidence for very early vigorous escape as in the first scenario.

There is some carbon fractionation on Mars (Webster et al. 2013) that is likely attributable to atmospheric escape, which suggests that carbon fractionation on Venus is possible if hydrogen escape took place early enough. Oxygen is expected to be mass fractionated on Venus if oxygen escaped with hydrogen, but the degree of O fractionation depends on how efficiently oxygen exchanged with crustal and mantle materials. Strong mass fractionation of oxygen on Venus would imply that the early water loss scenario is correct and that there was relatively little exchange with the crust and mantle; in contrast, a finding of weak fractionation would point to an increased likelihood of interaction between the mantle and the atmosphere (or hydrosphere) in Venus' past, and that the sink of oxygen from water was the mantle as proposed by Gillman et al. (2009) and Hamano et al. (2013). Finally, the present-day abundance of sulfur in the Venus atmosphere is tightly bound up in the planet's history of water, both chemically and as a driver of climate through the partnership between these materials in making sulfuric acid aerosols. The hygroscopic interactions between water and sulfur play a key role in limiting hydrogen escape from Venus today to very low rates, and are probably a major determinant of the D/H ratio of the hydrogen that does escape. Thus, a proper understanding of D/H fractionation in the Venus atmosphere today must take sulfur into account.

As the water loss history of Venus represents one of the most critical aspects of the planet's geological evolution , investigations that provide measurements with which to test the various models of early or later water loss scenarios remain a high priority for future mission development and design. Ascertaining the longevity of surface water on early Venus will provide important insights on general climate and habitability evolution in exoplanetary systems.

## 4. The Geological Enigma

Venus also presents several conundra in relation to its current state and past geologic and geodynamic evolution that are very relevant to its current atmospheric state. Prior to detailed spacecraft exploration, the similar size and density of Venus and Earth, and their proximity in the Solar System, suggested the possibility of a similar geologic evolution, but a divergent atmospheric evolution. Early Pioneer-Venus topography revealed "continent-like" highlands and linear lowlands (Pettengill et al. 1980), raising some key questions as to the geodynamic state of Venus. Were these configurations evidence of active plate tectonics and continental drift on Venus? Was Venus geologically and geodynamically Earth-like, but shrouded in an atmosphere dominated by a thermally and compositionally different atmosphere? Ensuing spacecraft radar imaging missions (the Soviet Venera 15 and 16, and the US NASA Magellan missions) revealed the global inventory of geological features, structures and units, the stratigraphic relationships and an interpretative sequence of events (Ivanov & Head, 2011). These investigations showed evidence for rift zones, folded mountain belts, vast volcanic plains, and many hundreds of volcanoes and circular to oval plume-like deformation features (coronae) (e.g., see summary in Head, 2014).

But what about the ages of different parts of the surface? The most surprising finding was that only ~1000 impact craters are preserved on the surface of Venus, yielding a remarkably young average age for the planet of ~750 Ma (Schaber et al., 1992; McKinnon et al., 1997). No heavily cratered (and so presumably older) areas analogous to the cratered terrains on the Moon, Mercury, and Mars were found. Importantly, the findings that the surface of Venus has an average age less than 20% of the total age of the planet, are *not* reflective of an amalgamation of very old and very young surfaces, such as the continents and ocean basins found on Earth, or the ancient highlands and the younger maria found on the Moon. The dearth of resolvable variability in crater areal density, and no sign of a braod range of crater degradation (as seen on the Moon, Mars and Mercury) suggested a similar age for all geological units. This inference, in turn, led to a view that the current surface of Venus evidently was produced within the past several hundred million years. It is also considered that the observed surface was possibly produced catastrophically, with minimal volcanic or tectonic resurfacing in the meantime. In this regard, the surface of Venus has evolved quite differently from Earth, the Moon, Mars, and Mercury. There have been numerous hypotheses that attempt to explain these surprising results, with limited progress toward generating a consensus (see summaries of these discussions in Bougher et al., 1997). Some workers called on geologically recent tectonic and volcanic catastrophic resurfacing. The mechanisms for such resurfacing include vertical crustal accretion and catastrophic overturn of a depleted mantle layer (Parmentier & Hess, 1992) and episodic occurrence of plate tectonics (Turcotte, 1993), followed by relative dormancy (Schaber et al., 1992). Others suggested that the surface evolution may have been triggered by a change in the mantle convection mechanism that was in turn related to the planetary thermal evolution, or alternatively caused by a transition from a mobile lid to a stagnant lid regime (Herrick, 1994).

Might any parts of the surface of Venus record an earlier history, possibly even dating back to a time when Venus could have possessed a much greater volume of water and so even been habitable? Several authors have argued that some locally elevated, highly deformed regions of the Venusian surface (called tesserae) may represent crustal materials that pre-date the surrounding terrain (Basilevsky & Head 1998; Head & Basilevsky 1998; Hansen et al. 1999; Gregg 2015). Such regions may record a distinct era from the Venusian past (Gilmore et al. 1997; Brown & Grimm, 1997; Gilmore & Head, 2018). Some studies have even suggested that tessera might have a lower emissivity than the surrounding dark basaltic plains, which could suggest materials with relatively high silica abundances that might be indicative of ancient crustal processes, possibly even continent building (Gilmore & Stein 2017). Alternatively, these emissivity variations could be due to grain size differences (Basilevsky et al. 2004; 2007) without any specific bearing on composition or formation. If tesserae or other areas of Venus' surface record an ancient history, targeted investigations of these regions might yield clues to the early state and evolution of the planet.

The nature of the geological and geodynamic evolution of the first 80% of Venus' history is currently unresolved and awaits new missions and investigations to assess a series of key unresolved questions[1]. It is clear, however, that the idea that the atmospheric evolution of Venus proceeded separately from the geological and geodynamic evolution is false (Phillips et al. 2001; Taylor & Grinspoon 2009; Lammer et al. 2018). Could Venus' current atmosphere have been produced by catastrophic degassing during geologically recent global volcanic resurfacing (Bullock & Grinspoon 1996; Solomon et al. 1999; Bullock & Grinspoon 2001)? Was a substantial amount of Venus' water inventory returned to the interior during short-lived global overturn events (Greenwood 2018) or longer-term plate tectonic-like recycling? Might Venus' atmospheric evolution be episodic, with multiple overturn and volcanic recycling events in its history (e.g. Strom et al. 1994; Moresi & Solomatov 1998; Armann & Tackley 2012)? Does the current surface temperature and environment influence the style of tectonics through deeply penetrating thermal effects (Ghail 2015; Platz et al. 2015)? Perhaps examples of these separate and divergent evolutionary paths are present in the cornucopia of discovered Venus-like exoplanets. Exploring this coupled geological-geophysical-atmosphere parameter space, by way of future missions to Venus, continued exoplanet surveys, and geodynamic and climate modeling, will help to crystallize an understanding of, and bring new insight to, the formative years of Venus as well as Earth.

---

[1] https://www.lpi.usra.edu/vexag/reports/

Although Venus accounts for 40% of the mass of terrestrial planets in our solar system, fundamental properties of its interior composition have been predicted (e.g. Zharkov, 1983), but not yet measured. As we expand the scope of planetary science to include those planets around other stars, , core-size and state, seismic velocity and density variations with depth, and thermal profiles, will provide us with critical benchmarks for testing geochemical and geodynamic models of terrestrial and exoplanet interiors in general. Furthermore, measurements of the relative abundances of Venus' refractory elements can greatly inform models of the degree of mixing of planetesimals within the critical zone of the protoplanetary disk where terrestrial planets formed. If the relative refractory element ratios found for the Venus surface are reflected in the size of the core, we gain a key benchmark for studies of how this and other planetary systems formed by constraining even this simple parameter for Venus. Such a finding, in turn, will greatly aid in our studies of exoplanets, where stellar composition may set the initial compositional gradient of planetesimals within the disk but for which the degree of mixing remains an elusive, underconstrained parameter.

## 5. A Plethora of Venus Analogs

The HZ boundaries for main-sequence stars have been previously studied utilizing a variety of climate models, such as those by Kasting et al. (1993), and more recently by Kopparapu et al. (2013, 2014). These HZ calculations provide a fundamental application in estimating the fraction of stars that harbor Earth-size planets in the HZ region, or eta-Earth. Much of the eta-Earth calculations utilize results from the *Kepler* mission since these data provide a uniformly derived sample of many terrestrial-size planets to which meaningful statistical analyses can be applied (Dressing & Charbonneau 2013, 2015; Kopparapu 2013; Petigura et al. 2013).

The transiting exoplanet detection method is strongly biased towards the discovery of planets that are relatively close to the host star (Kane & von Braun 2008). Furthermore, shorter orbital periods result in enhanced signal-to-noise (S/N) ratios of transit signatures because of the larger number of observed transits within an observatuonal window . Consequencely, data from the *Kepler* mission has resulted in detected planets that are preferentially interior to the HZ, the insolation flux of which means that they are more likely to be potential analogs to Venus rather than Earth . Since the prospect of a divergence in the evolution of the Earth and Venusian atmospheres is an important factor in understanding the habitability of Earth, the occurrence rate of Venus analogs (i.e., eta-Venus) is also an important parameter to quantify.

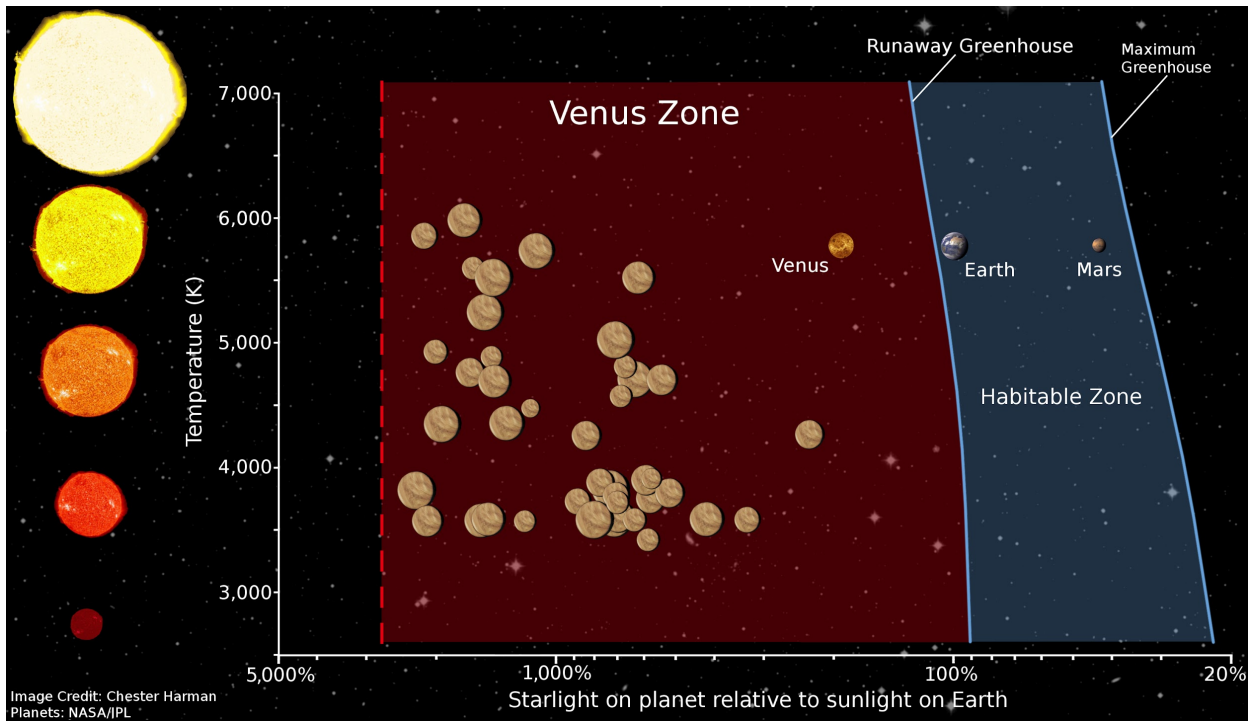

*Figure 2: The extent of the Venus Zone as a function of host star temperature and incident flux, where solar system planets and Kepler candidates in the terrestrial regime are shown. Credit: Chester Harman.*

Kane et al. (2014) defined the "Venus Zone" (VZ) as a target selection tool to identify terrestrial planets, as a function of instellation flux, where the atmosphere could potentially be pushed into a runaway greenhouse producing surface conditions similar to those at Venus. Figure 2 shows the VZ (red) and HZ (blue) for stars of different temperatures. The outer boundary of the VZ (inner blue line) is the "Runaway Greenhouse" line, calculated using climate models of Earth's atmosphere (Kane et al. 2014, Kopparapu et al. 2013, 2014). The inner boundary of the VZ (red dashed line) is estimated to be where stellar radiation from the star would cause complete atmospheric erosion (Zahnle & Catling 2017). The pictures of Venus shown in this region represent planet candidates detected by *Kepler, where the size of the pictures are scaled to the size of the detected planets*. Kane et al. (2014) calculated occurrence rates of potential Venus analogs by examining the *Kepler* exoplanet candidates that were discovered within different ranges of orbital periods, and comparing those numbers with the expected values based on the known demographics of exoplanetary systems and accounting for the bias of the transit method toward shorter orbital periods, described above. These calculations yielded occurrence rates of VZ terrestrial planets of 32% for low-mass stars (M dwarfs) and 45% for Sun-like stars (K and G dwarfs). However, note that, as for the HZ, the boundaries of the VZ should be considered a testable hypothesis since runaway greenhouse could occur beyond the calculated boundary (Hamano et al. 2013; Foley 2015).

There are now numerous examples of terrestrial planets whose orbits lie interior to the runaway greenhouse limit of their host stars. The top-down view of the orbital architectures for two of these systems are shown in Figure 3; the K2-3 (left, Crossfield et al. 2015) and TRAPPIST-1 (right, Gillon et al. 2017) planetary systems, where the green regions indicate the extent of the HZ (Kane & Gelino 2012). The light green region is referred to as the "conservative" HZ, bounded by the runaway greenhouse and maximum greenhouse limts shown in Figure 2. The dark green regions represent the "optimistic" HZ, and are calculated based on assumptions regarding the possible past surface liquid water history of Venus and Mars (Kane et al. 2016). The scale of the figure along one edge is indicated above each system. For both systems, planet d lies near the inner edge of the optimistic HZ, which is where the VZ overlaps with the HZ and is calculated under an optimistic assumption that Venus could have maintained liquid surface water until as recently as 1 Gya (Kopparapu et al. 2013). Both of these planets are also terrestrial and are prime candidates for further study as potential Venus analogs.

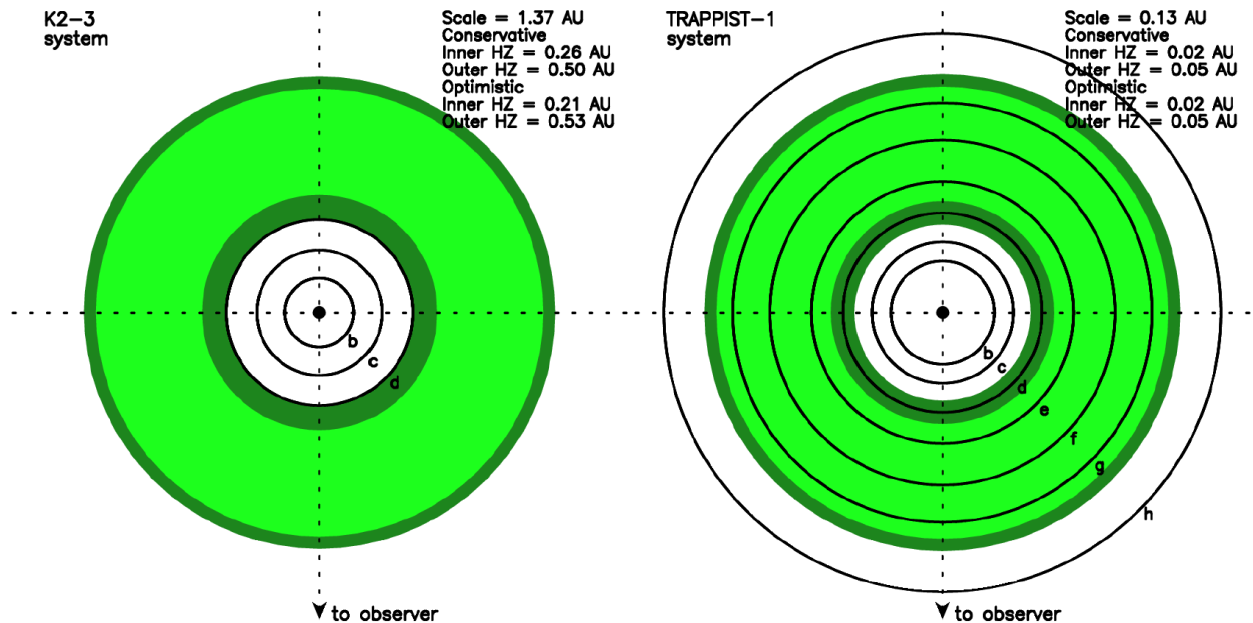

*Figure 3: A top-down view of the K2-3 and TRAPPIST-1 planetary systems, showing the orbits of the planets and the extent of the HZ (green) annd VZ (dark green).*

The occurrence rate of Venus analogs will continue to be relevant in the current era of the *Transiting Exoplanet Survey Satellite (TESS)* mission, as hundreds of terrestrial planets orbiting bright host stars are expected to be detected (Sullivan et al. 2015; Huang et al. 2018). Simulations by Huang et al. (2018) predicted that *TESS* will discover ~100 planets smaller than 1.25 Earth radii during the primary mission, of which ~30 will orbit stars brighter than a *TESS* magnitude of 10, making the stars amongst the brightest of those hosting transiting terrestrial planets. The discoveries of *TESS* will thus provide key opportunities for transmission spectroscopy follow-up observations during planetary transit using *JWST*, amongst other facilities (Seager & Sasselov 2000; Kempton et al. 2018), and will be used to investigate upper atmosphere compositions. Such observations that are capable of characterizing the atmospheric compositions of terrestrial planets will need to face the challenge of distinguishing between possible Venus and Earth-like surface conditions by combining observations of identifying relative amounts of atmospheric greenhouse gases with geological models that accounts for those abundances (Schaefer & Fegley 2011; Ehrenreich et al. 2012; Kane et al. 2018). Apart from wavelength coverage and signal-to-noise considerations, the challenge of correct interpretation of transmission spectra arises from such modeling aspects as atmospheric opacity as a function of scale height and the degeneracy between models that can distinguish between runaway greenhouse and temperate surface conditions (Robinson 2017). Overcoming these challenges will be combined with a simultaneous statistical analysis of potential Venus analogs and their occurrence rates, leading to a quantitative assessment of the primary contributors toward the emergence of runaway greenhouse atmospheres and thus allowing us to decode why the atmosphere of Venus may have so radically diverged from its sibling planet, Earth.

## 6. Understanding the Extrema of Habitability

Many significant questions regarding the current state of Venus remain, pointing to major gaps in our understanding of the evolution of terrestrial planets, including the future evolution of Earth. Major outstanding questions include:

- What is the interior structure and bulk composition of Venus? How much does it differ from that of Earth and of the Sun?
- Did Venus have a habitable period (e.g. Way et al. 2016)? That is, did Venus ever cool from a syn-accretionary runaway greenhouse (Hamano et al. 2013)? If Venus had a habitable period, how long did it last? Are there possible habitable locations within the cloud layers where temperate conditions exist (Limaye et al. 2018)?
- Are there any remnants of ancient crustal materials on the surface formed from silica-rich minerals (Hashimoto & Sugita 2003)?
- Where did the water go? Was hydrogen loss and abiotic oxygen production prevalent, or did surface hydration dominate?
- What has been the history of tectonics activity and deformations, volatile cycling, and volcanic resurfacing (Ivanov & Head 2011) on Venus? Was the delivery of volatiles to the atmosphere gradual, episodic, or catastrophic? When did Venus enter its present stagnant-lid regime? Does any subduction occur today (Davaille et al. 2017; Smrekar et al. 2018)?
- What is the detailed composition, structure, and chemical reactions that exist within the Venusian middle and deep atmosphere and how does the atmosphere interact with the surface?

A major focus of current and future exoplanet science is the measurement and modeling of terrestrial atmospheres. Interpretation of these data are sensitive to the composition, chemistry, and dynamics of the deeper atmosphere which is largely opaque at most wavelengths. It is therefore imperative to obtain additional in situ data for terrestrial atmospheres within our solar system, particularly for a diverse range of atmospheric chemistries. Figure 4 summarizes some of the outstanding questions that remain to be addressed for Venus, encompassing the various layers of the Venusian atmosphere as well as the surface and interior, described in more detail by Taylor et al. (2018). The lack of in situ data to constrain models of the Venus atmosphere, and the difficulty in building robust models with those data currently available , substantially inhibit in efforts to effectively model the surface environments of terrestrial planets outside of the solar system. Moreover, a detailed knowledge of Venus' atmosphere will provide us with a benchmark for modeling the coevolution of the planet's surface and interior and will further aid in our ability to interpolate and extrapolate similar processes in exoplanet models.

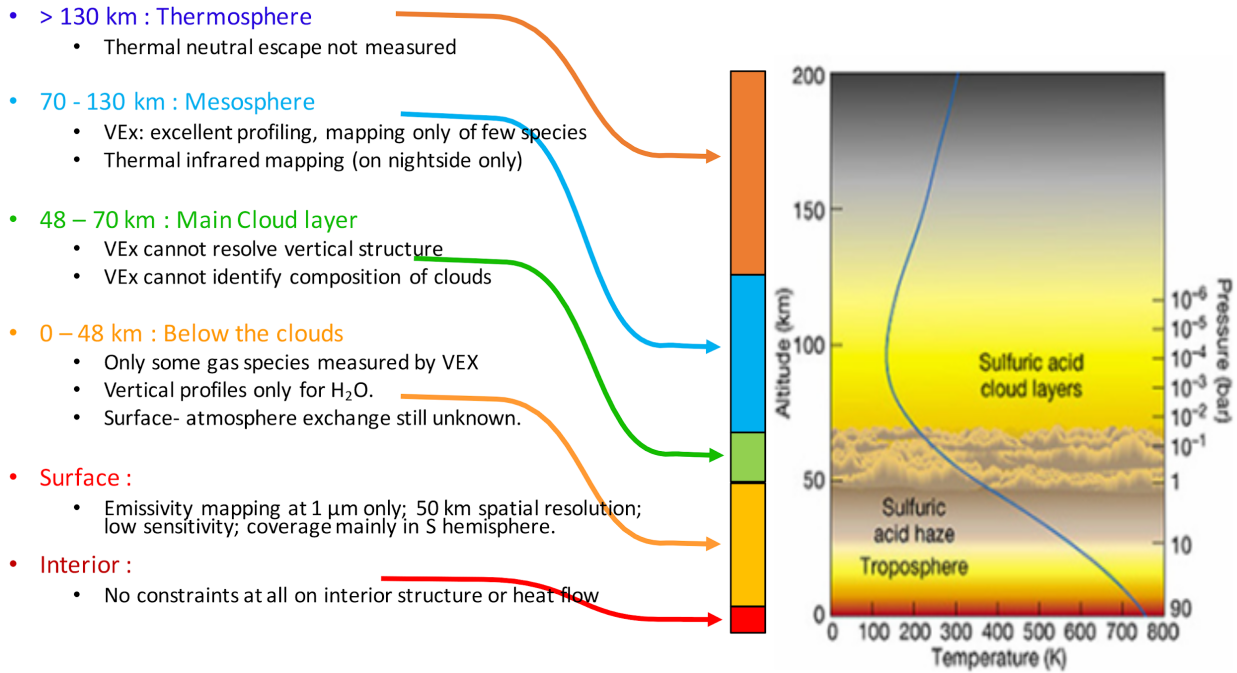

- \> 130 km : Thermosphere
  - Thermal neutral escape not measured

- 70 - 130 km : Mesosphere
  - VEx: excellent profiling, mapping only of few species
  - Thermal infrared mapping (on nightside only)

- 48 – 70 km : Main Cloud layer
  - VEx cannot resolve vertical structure
  - VEx cannot identify composition of clouds

- 0 – 48 km : Below the clouds
  - Only some gas species measured by VEX
  - Vertical profiles only for $H_2O$.
  - Surface- atmosphere exchange still unknown.

- Surface :
  - Emissivity mapping at 1 µm only; 50 km spatial resolution; low sensitivity; coverage mainly in S hemisphere.

- Interior :
  - No constraints at all on interior structure or heat flow

*Figure 4: A summary of some outstanding questions regarding the atmosphere and surface of Venus (Taylor et al. 2018).*

## 7. Conclusions

The only in situ terrestrial planetary data available to us are here in our solar system, and there are many opportunities to study exoplanet analogs from our terrestrial body inventory. Since efforts towards the detection and characterization of exoplanets are focussed on Earth-size planets, Venus is an ideal and accessible exoplanet laboratory. Data from Venus have wide-reaching consequences for studying exoplanets, and may be applied to modeling planetary atmospheres, surfaces, interiors, and geological processes that contribute to detectable atmospheric signatures. The next greatest advances in studies of Venus will come from improved answers to the top-level questions described in Section 6; in particular, the finding of evidence for previously temperate conditions on Venus would significantly enhance studies of habitability and our understanding of the prevalence of life in the universe.

Atmospheric modeling of exoplanets is also of critical importance and improved measurements of pressure, temperature, composition, and dynamics of the Venusian atmosphere as a function of latitude and altitude would aid enormously in our ability to interpret and model exoplanetary atmospheres. In particular, new direct measurements of D/H ratios within and below the clouds are needed to better constrain the historical volume of water on Venus.

Combined with D/H, isotopic measurements in the atmosphere would yield insights into the origins and fate of the Venusian atmosphere. Further measurements of the Venusian deep atmosphere will allow a detailed study of the atmospheric chemistry that occurs at very high temperature and pressures. Such measurements are important for exoplanet atmospheric studies because the deep atmosphere of exoplanets will be inferred from models that use data of the upper atmosphere obtained via transmission spectroscopy (Hu et al. 2012; Forget & Leconte 2014).

Detailed knowledge of the Venusian interior also plays an important role in our ability to construct robust models of exoplanetary interiors. Current interior models of exoplanets are based upon limited solar system data and phase transition diagrams combined with mass/radius measurements and extraction of stellar abundance information (Valencia et al. 2007; Dorn et al. 2015; Hinkel & Unterborn 2018; Wang et al. 2019). Such interior modeling efforts would benefit enormously from additional data of the interior of Venus since the planet, along with Earth, best represent the terrestrial planets that are accessible via current exoplanet detection methods. Specifically, the most valuable interior data will come from measurements that refine Venus' moment of inertia and allow for the determination of the planet's geologic evolution, its current level of activity, and indications of key geodynamic changes (e.g., tectonic and thermal regime) with time. These fundamental measurements would stimulate progress in addressing the key questions described in Section 6 on multiple fronts, and vastly improve our understanding of both modern Venus and its pathway to that modern state.

Ultimately, Venus is an exoplanet laboratory next door that presents an opportunity to conduct a detailed study of planetary atmospheres and the evolution of habitability that will never be available to us elsewhere. The considerable number of unanswered, major questions regarding Venus and their profound bearing on the correct interpretation of atmospheric data and the connection to the geophysics of the planet, means that we must recognize the consequential limitations of our ability at present to reliably infer the surface conditions of exoplanets for which data will always be several orders of magnitude less accessible. Importantly, and despite its current surface environment, Venus has a vital story to tell regarding the evolution of a habitable planet, from starting conditions that may have been similar to Earth, through a period of temperate climates, to an eventual fall into post-runaway greenhouse calamity. It is critical, now more than ever, that we consider carefully that story.

**Acknowledgements**
The authors would like to thank the referees, Paul Byrne and Richard Ghail, whose detailed comments greatly improved the quality of the manuscript. This research has made use of the following archives: the Habitable Zone Gallery at hzgallery.org and the NASA Exoplanet Archive, which is operated by the California Institute of Technology, under contract with the National Aeronautics and Space Administration under the Exoplanet Exploration Program. The results reported herein benefited from collaborations and/or information exchange within NASA's Nexus for Exoplanet System Science (NExSS) research coordination network